\documentclass[page-classic]{epl2} 

\usepackage{graphicx} 

\title{Thermal conductance of the coupled-rotator chain: 
Influence of temperature and size}
\shorttitle{Thermal conductances of the 1D coupled rotator lattice}

\author{Yunyun Li\inst{1,2,3}, Nianbei Li\inst{1,2,3}, Ugur Tirnakli\inst{4,5}, 
   Baowen Li\inst{6}, Constantino Tsallis\inst{5,7}}
\shortauthor{Y. Li  \etal}

\institute{
\inst{1} Center for Phononics and Thermal Energy Science and 
School of Physics Science and Engineering, 
Tongji University, 200092 Shanghai, China\\
\inst{2}China-EU Joint Lab for Nanophononics, Tongji University, Shanghai 200092, China \\
\inst{3}Shanghai Key Laboratory of Special Artificial Microstructure Materials and Technology, School of Physics Science and Engineering, Tongji University, Shanghai 200092, China\\
\inst{4} Department of Physics, Faculty of Science, Ege University, 
35100 Izmir, Turkey\\
\inst{5} Centro Brasileiro de Pesquisas Fisicas and National Institute 
for Science and Technology of Complex Systems, 
Rua Dr. Xavier Sigaud 150, 22290-180 Rio de Janeiro, RJ, Brazil\\
\inst{6} Department of Mechanical Engineering, University of Colorado Boulder, 
CO 80309, USA\\
\inst{7} Santa Fe Institute, 1399 Hyde Park Road, Santa Fe, New Mexico 87501, USA}

\abstract{
Thermal conductance of a homogeneous 1D nonlinear lattice system with neareast neighbor 
interactions has recently been computationally studied in detail by Li et al [Eur. Phys. J. B {\bf88}, 
182 (2015)], where its power-law dependence on temperature $T$ for high temperatures is 
shown. Here, we address its entire temperature dependence, in addition to its dependence on 
the size $N$ of the system. We obtain a neat data collapse for arbitrary temperatures and 
system sizes, and numerically show that the thermal conductance curve is quite satisfactorily 
described by a fat-tailed $q$-Gaussian dependence on $TN^{1/3}$ with $q \simeq 1.55$. 
Consequently, its $T \to\infty$ asymptotic behavior is given by 
$T^{-\alpha}$ with $\alpha=2/(q-1) \simeq 3.64$. 
}

\begin{document}

\maketitle

\section{Introduction}
The breakdown of Fourier's law in low-dimensional lattices has attracted much attention in 
recent years due to its fundamental importance within non-equilibrium thermodynamics and 
statistical mechanics \cite{Lepri,Dhar,Liu,Casati,Olivares2016,Bagchi2017}. In the 1D Fermi-Pasta-Ulam (FPU-$\beta$) 
lattice, the thermal conductivities $\kappa$ diverge with system sizes $N$ as 
$\kappa\propto N^{\gamma}$, where $0<\gamma<1$; consequently its thermal conductance 
$\sigma \equiv \kappa/N$ vanishes in the $N\to\infty$ limit.  However, there is still no clear 
conclusion about the physical ingredient responsible for this kind of anomalous heat conduction. 
It is believed that momentum conservation is the crucial reason for the anomalous heat 
conduction \cite{Li2,Prosen,Narayan}, but normal heat conduction has been found in 1D 
coupled rotator lattice, which also is a momentum-conserved system \cite{Vassalli,Gendelman}.

Unlike the FPU-$\beta$-like lattices, the 1D coupled rotator lattice has periodic interatomic 
potential which is finite. As a result, the energy diffusion as well as the momentum diffusion are 
normal \cite{NJP}. In order to understand the effect of this finite interatomic potential, previous 
works focus on the temperature dependence of the thermal conductivities in 1D coupled rotator 
lattice \cite{Vassalli,Gendelman}. In both works, the thermal conductivity was proposed to have 
an exponential dependence on temperature. But it is argued that 
$\kappa(T) \propto e^{\Delta V/T}$ where $\Delta V$ is proportional to the potential barrier 
height in \cite{Vassalli}, while $\kappa(T) \propto e^{-T/A}$ with $A$ a fitting parameter 
in \cite{Gendelman}. It has only recently been found that the temperature dependence in 1D 
coupled rotator lattice follows a power-law behavior on temperature as 
$\kappa(T)\propto T^{-\alpha}$ with $\alpha \simeq 3.2$ for intermediate 
temperatures \cite{EPJB}. Interestingly enough, this power-law dependence is 
qualitatively consistent with the theoretical prediction for the Chirikov standard map which is a 
single rotator model \cite{Chirikov1979pr,MacKay1984pd}.

On the other hand, the standard map, as well as several other dynamical complex systems, 
has recently been shown \cite{TirnakliBorges} to present non-Gaussian probability distributions 
for the sum of its position random variable. These distributions are  approached extremely well 
by the $q$-Gaussian defined as 

\begin{equation}
 \label{q-Gauss}
P_q(x)=A_q \, e_q^{-B_qx^2} \equiv \frac{A_q}{[1+(q-1)B_q\, x^2]^{1/(q-1)}},
\end{equation}
where $A_q$ is the normalization factor, $B_q >0$ is a parameter which characterizes
the width of the distribution, and the index $q \ge 1$ \cite{Tsallis1988,prato-tsallis-1999,Tsallis2009}. In the $q\to 1$ limit, this expression recovers the 
standard Gaussian distribution. This family of distributions optimizes the nonadditive entropy $S_q =k \frac{1-\int dx\,[p(x)]^q}{q-1}$ with $S_1=S_{BG} \equiv -k \int dx\, p(x) \ln p(x)$ under appropriate constraints, where $k$ is Boltzmann constant, $p(x)$ is the probability distribution, and BG stands for Boltzmann-Gibbs. \\

\section{Model and Method}
In this letter, we study a homogeneous 1D nonlinear lattice system with nearest neighbor 
interactions and try to see whether some of its properties  also are consistent with 
$q$-Gaussians. For this lattice system, the Hamiltonian with the corresponding dimensionless 
units can be written in the general form

\begin{equation}
\label{ham}
H=\sum_{i=1}^N \left[\frac{p_{i}^2}{2} + V(q_{i+1}-q_{i}) + U(q_{i})\right] \;
\label{Hamiltonian1}
\end{equation}
where $p_i$ denotes the momentum for the $i$-th rotator. The set $q_i$ are the displacements 
from the equilibrium position for the $i$-th rotator; $V(q_{i+1}-q_{i})$ is the interaction potential 
between neighboring sites $i$ and $i+1$, and $U(q_{i})$ is the on-site potential, representing 
the interaction with the substrate. To focus on the momentum-conserving system, we set 
$U=0$. The potential we employ is in the form 

\begin{equation}
\label{V}
V(x) = V_0(1-\cos x) \;
\label{Hamiltonian2}
\end{equation}
where $V_0$ is the interaction strength (without loss of generality we set $V_0=1$).

In our simulation, a Langevin form of heat bath is used. For the chain with $N$ particles, only 
the first and last particles are coupled to the heat bath, with the temperature $T_L$ and $T_R$, 
respectively. The dynamics equations of the motion are read as

\begin{eqnarray}
\label{Dynamics}
\dot{q_i}&=& p_i,  i = 1,2, 3, ..., N,\\ \nonumber
\dot{p_i}&=& F(q_{i}-q_{i-1})+F(q_{i+1}-q_{i}), \mathrm{i = 2, 3, ..., N-1},\\ \nonumber
\dot{p_1}&=& F(q_{i})+F(q_{i+1}-q_{i})-\gamma p_1+\xi_1, \\ \nonumber
\dot{p_N}&=& F(q_{i}-q_{i-1})+F(-q_{i})-\gamma p_N+\xi_N \nonumber
\end{eqnarray}
where $F(x)=-\partial{V(x)}/\partial{q_i}$ and $\gamma$ is the friction coefficient; 
$\xi_1$, $\xi_N$ are Gaussian white noise with zero mean $<\xi_1(t)>=0$ and $<\xi_N(t)>=0$. 
The correlation function is given by 

\begin{eqnarray}
\label{random}
<\xi_1(t)\xi_1(t^{\prime})>=2\gamma T_{L} \delta(t-t^{\prime}), \\ \nonumber
<\xi_N(t)\xi_N(t^{\prime})>=2\gamma T_{R} \delta(t-t^{\prime}). \nonumber
\end{eqnarray}
For simplicity, the temperatures are set as $T_{L/R} = T_0 (1 \pm \Delta)$, where $T_0$ is 
the average temperature and $\Delta$ is the temperature difference. Throughout our numerical 
simulations, $\Delta=0.1$ is restricted to the small perturbation regime and $\gamma=1$ is 
fixed. The evolution of dynamics (Eqs. (\ref{Dynamics})) is integrated by the Verlet velocity  
algorithm and the time step $\Delta t=0.01$, which is small enough \cite{verlet}. 
All the results are analyzed for the time scale $10^7-10^8$, after the system release to the 
steady state. \\

\begin{figure}
\includegraphics[width=0.95\columnwidth]{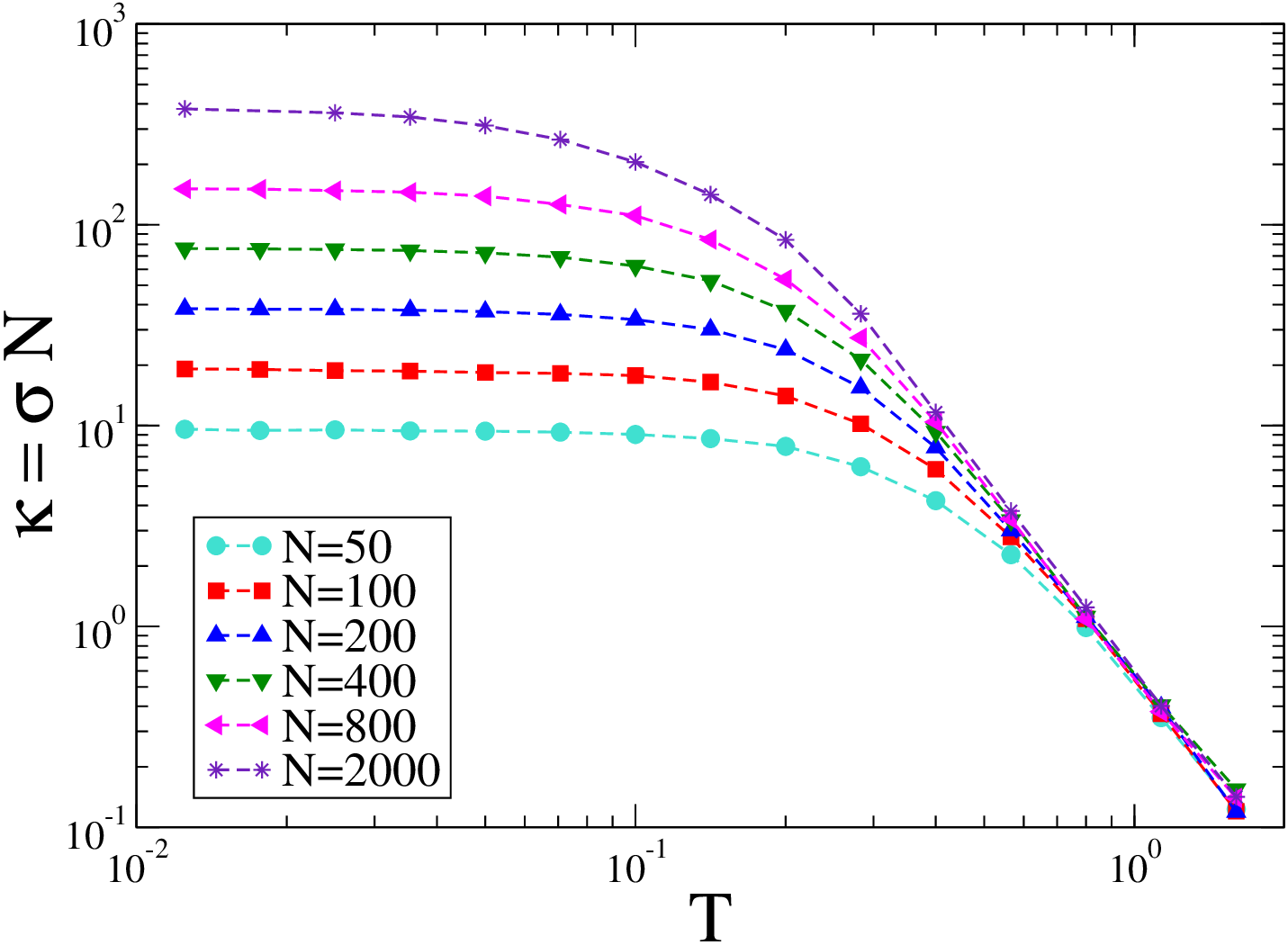}
\includegraphics[width=0.95\columnwidth]{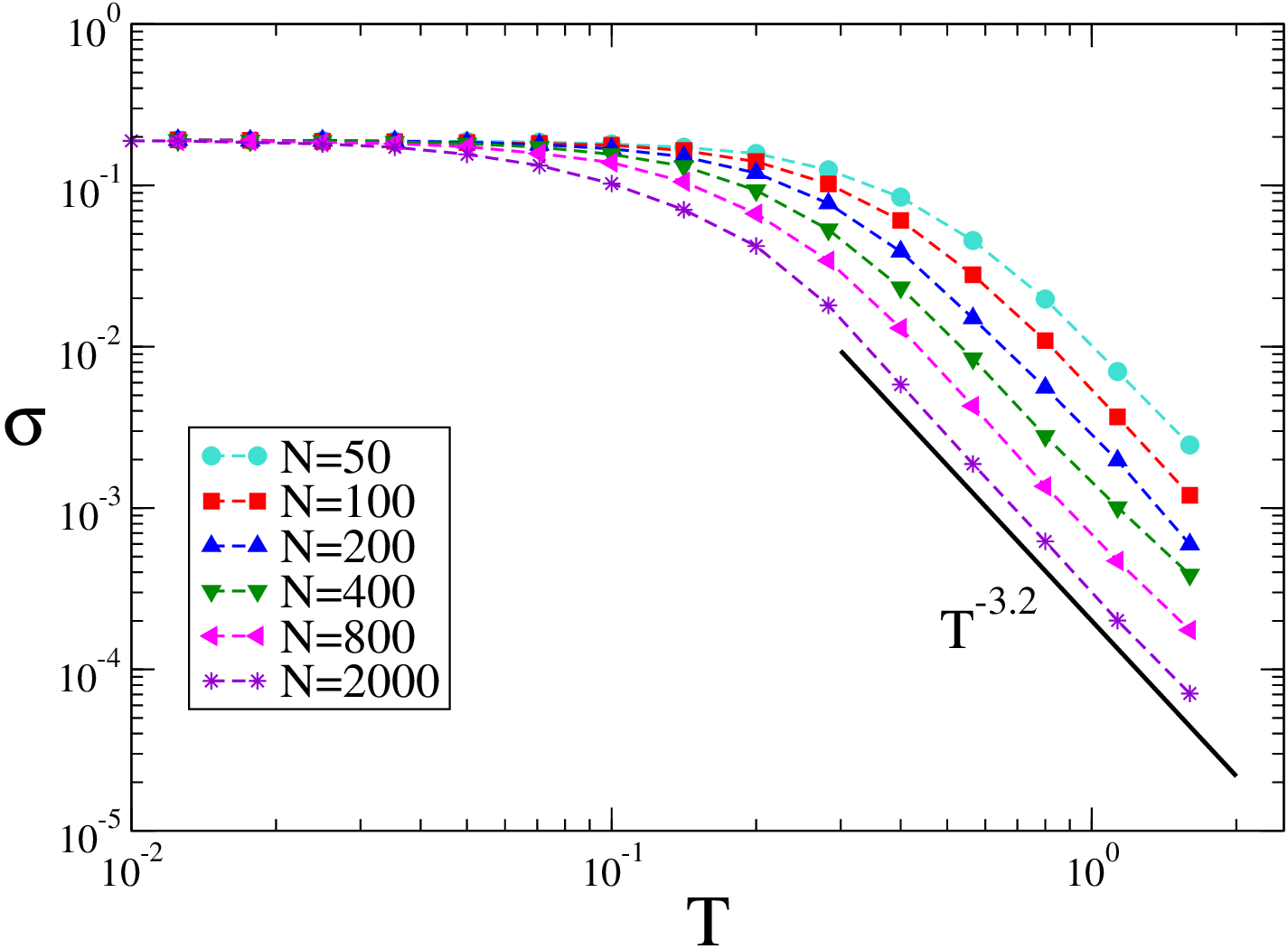}
\caption{\label{Q1} (Color online) Heat flow at stationary state for typical lattice sizes. 
{\it Top:} Thermal conductivity $\kappa \equiv \sigma N$ (notice that data collapse occurs for the high-temperature region); {\it Bottom:} Thermal conductance $\sigma$ (notice that data collapse occurs for the low-temperature region).
Different colors correspond to the lattice lengths $N=50$, 100, 200, 400, 800 and 2000. 
The slope $-3.2$ indicated in \cite{EPJB} is shown here for comparison. 
The dashed curves are guides to the eye.}
\label{original}
\end{figure}

\begin{figure}
\includegraphics[width=0.95\columnwidth]{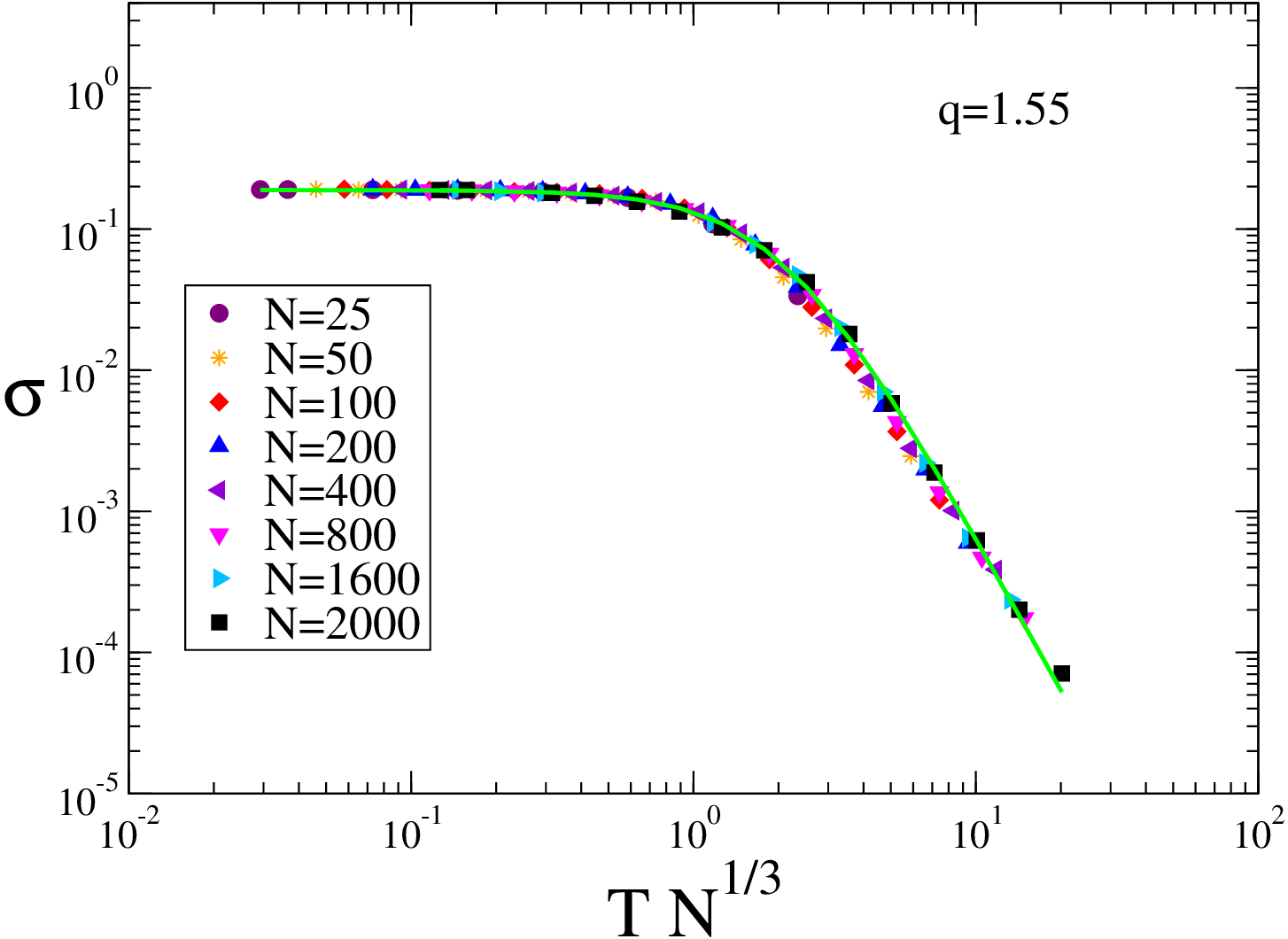}
\caption{\label{Q2} (Color online) Data collapse for lattice sizes going from $N=25$ to 
$N=2000$. The continuous curve (green line) corresponds to 
$\sigma=A_q e_q^{-B_q\,(TN^{1/3})^2}$ with $(q,B_q,A_q)=(1.55,0.40,0.189)$. 
The asymptotic slope is given by $2/(1-q) \simeq - 3.64$, in contrast with the intermediate 
slope $-3.2$ indicated in \cite{EPJB} (see Fig. \ref{original}).}
\label{collapse}
\end{figure}

\section{Results and Discussion}
The thermal conductivity $\kappa$ is characterized by

\begin{equation}
\label{kappa}
\kappa(T)= \frac{JN}{T_L-T_{R}} \;
\end{equation}
where $J=< J_i >$ is the average heat flux along the lattice and $J_i$ is the local heat flux. 
As already mentioned, the thermal conductance is defined as $\sigma \equiv \kappa/N$. 
The temperature dependence of thermal conductance is given in Fig.\ \ref{Q1} for six different 
lattice sizes. The asymptotic power-law behavior is evident with an exponent $-3.2$. 
One can easily obtain a clear data collapse as shown in Fig.\ \ref{Q2}. It is 
evident that the temperature dependence of thermal conductance can be satisfactorily 
approached by a $q$-Gaussian with $q=1.55$. \\

In conclusion, we have numerically determined that the thermal conductance for the classical
one-dimensional first-neighbor coupled planar-rotator (or {\it inertial XY ferromagnetic}) chain (Eqs. (\ref{Hamiltonian1}) and (\ref{Hamiltonian2}), with vanishing on-site potential $U(q_i)$ and unit potential strength $V_0$)
is amazingly well described by the 
$q$-Gaussian $\sigma\propto e_{1.55}^{-0.40\,(TN^{1/3})^2}$ for wide ranges of temperatures 
$T$ and lattice sizes $N$. This result implies that, in the $(T N^{1/3}) \to\infty$ limit, 
we asymptotically expect $\sigma \propto [TN^{1/3}]^{-\alpha}$ with 
$\alpha = 2/(q-1) \simeq 3.64$, close though different from the value 3.2 determined 
in \cite{EPJB} for intermediate temperatures. At thermal equilibrium (i.e., for $T_L=T_R$), 
it is clear that the present short-range-interacting model follows Boltzmann-Gibbs statistical 
mechanics. Why then, in the nonequilibrium stationary state characterized by $T_L \ne T_R$, 
such a strong suggestion of $q$-statistics emerges? This remains as a highly interesting and 
certainly intriguing open question,
somewhat reminiscent of the aging and related phenomena in various systems: see for example \cite{cugliandolo1995} (its Fig. 1), \cite{Stariolo2003}, and \cite{Fyodorov2015} (its Fig. 1, for instance); see also \cite{Doye} (its Fig. 3). It is of course not excluded that, due to the permanent unidirectional heat 
flow, the phase space of the chain is visited in an incomplete manner. Further understanding 
would of course be very welcome.

\acknowledgments
Y. L., N. L. and B. L. are supported by the NSF China with grant No. 11334007. U. T. is a member of the Science Academy, Istanbul, Turkey. 
Two of us (U. T. and C. T.) acknowledge partial financial support from CNPq, Capes and 
Faperj (Brazilian agencies), as well as from the John Templeton Foundation (USA).

\end{document}